\documentclass[12pt,preprint]{aastex}

\usepackage{epsfig}
                                                                                
\def\vkm{km s$^{-1}$}

\def\degree{$^\circ$}
\def\arcs#1{$#1''$}
\def\arcsa#1#2{$#1^{\prime\prime}_{^\textrm{.}}#2$}
\def\arcsaq#1#2{#1^{\prime\prime}_{^\textrm{.}}#2}
\def\smassrate{$M_\odot$ yr$^{-1}$}
\def\solarmass{$M_\odot$}

\def\solarlum{$L_\odot$}

\def\Jupmass{$M_\textrm{\scriptsize Jup}$}

\def\Jyb{Jy beam$^{-1}$}
\def\mJyb{mJy beam$^{-1}$}

\def\Jybk{Jy beam$^{-1}$ km s$^{-1}$}

\def\tlabel#1{(\textit{#1})}
\def\vrot{v_\textrm{rot}}

\def\cmc{cm$^{-3}$}
\def\cms{cm$^{-2}$}
\def\micron{$\mu$m}
                                                                                
\def\ra#1#2#3#4{#1^\mathrm{h} #2^\mathrm{m} #3^\mathrm{s}_{^\textrm{.}} #4}
\def\dec#1#2#3#4{#1\degr #2\arcmin #3^{\prime\prime}_{^\textrm{.}}#4}

\def\Lbol{L_\textrm{\scriptsize bol}}
\def\SOt{$N_J=8_9-7_8$}
\def\H2{H$_2$}
\def\N2HP{N$_2$H$^+$}

\def\NH3{NH$_3$}

\def\SOt{$N_J=8_9-7_8$}

\def\HCOP{HCO$^+$}
\def\aHCOP{H$^{13}$CO$^+$}
                                                                                
\def\putfigs#1#2#3{\epsfig{scale=#1,angle=#2,figure=#3}}
\def\putfig#1#2#3{}
\def\leftblank#1{}

\begin{document}

\title{Rotation and Outflow motions in the very low-mass Class 0 protostellar system HH 211
at subarcsecond resolution}
\author{Chin-Fei Lee\altaffilmark{1},
Naomi Hirano\altaffilmark{1}, Aina Palau\altaffilmark{2},
Paul T.P. Ho\altaffilmark{1,3}, 
Tyler L. Bourke\altaffilmark{3}, 
Qizhou Zhang\altaffilmark{3}, and
Hsien Shang\altaffilmark{1} 
}
\altaffiltext{1}{Academia Sinica Institute of Astronomy and Astrophysics,
P.O. Box 23-141, Taipei 106, Taiwan; cflee@asiaa.sinica.edu.tw}
\altaffiltext{2}{
Laboratorio de Astrof\'{\i}ica Estelar y Exoplanetas, Centro de                                          
Astrobiolog\'{\i}a (INTA-CSIC),                                                                         
LAEFF campus, P.O. Box 78, E-28691 Villanueva de la Ca\~nada (Madrid), Spain}
\altaffiltext{3}{Harvard-Smithsonian Center for Astrophysics, 60 Garden
Street, Cambridge, MA 02138}

\begin{abstract} 
HH 211 is a nearby young protostellar system with a highly
collimated jet. We have mapped it in 352 GHz continuum, SiO ($J=8-7$), and
\HCOP{} ($J=4-3$) emission at up to $\sim$ \arcsa{0}{2} resolution with the
Submillimeter Array (SMA). The continuum source is now resolved into two
sources, SMM1 and SMM2, with a separation of $\sim$ 84 AU. 
SMM1 is seen at the center of the jet, probably tracing a (inner) dusty disk
around the protostar driving the jet. SMM2 is seen to the
southwest of SMM1 and may trace an envelope-disk around a small binary
companion. A flattened envelope-disk is seen in \HCOP{} around SMM1
with a radius of $\sim$ 80 AU perpendicular to the jet axis.
Its velocity structure is consistent with a
rotation motion and can be fitted with
a Keplerian law that yields a mass of $\sim$ 50$\pm$15
\Jupmass{} (a mass of a brown dwarf) for the protostar. Thus, the protostar
could be the lowest mass source known to have a collimated jet and a
rotating flattened envelope-disk. A small-scale ($\sim$ 200 AU) low-speed ($\sim$ 2
\vkm{}) outflow is seen in \HCOP{} around the jet axis extending from the
envelope-disk. It seems to rotate
in the same direction as the envelope-disk and may carry away
part of the angular momentum from the envelope-disk. The jet is seen in SiO
close to $\sim$ 100 AU from SMM1.
It is seen with a ``C-shaped" bending. It has
a transverse width of $\lesssim$ 40 AU and a velocity of $\sim$ 170$\pm$60
\vkm{}. A possible velocity gradient is seen consistently across its innermost pair of
knots, with $\sim$ 0.5 \vkm{} at $\sim$ 10 AU, consistent with the sense 
of rotation of the envelope-disk. If this gradient is
an upper limit of the true rotational gradient of the jet,
then the jet carries away a very small amount of angular
momentum of $\lesssim$ 5 AU \vkm{} and thus must be launched from the
very inner edge of the disk near the corotation radius.
\end{abstract}


\section{Introduction}

Stars are formed inside molecular cloud cores by means of gravitational
collapse. The details of the process, however, are complicated by the
presence of magnetic fields and angular momentum. In particular, excess angular
momentum will have to be removed in order for stars to form. 
Part of it may be carried away by the jets
that are believed to be launched
from accretion disks around protostars.
Measurements of angular
momentum have been reported for the jets in various evolutionary phases from Class 0
\citep{Lee2008} to Class I \citep{Chrysostomou2008},
and to T-Tauri phase \citep{Coffey2007}. Most of these measurements,
however, are based on shock emission (e.g., SiO, \H2{}, and [OI]) and could
be uncertain. 
Without resolving the shock structures and kinematics,
the measurements can be
significantly affected by internal (bow) shock interactions 
\citep{Codella2007,Lee2008}. In addition to the jets,
low-speed molecular outflows have also been seen along the jet axis.
Some of them seem to be rotating and may also
carry away part of the angular momentum \citep{Lee2007HH212,Launhardt2008}.

This paper is a follow-up to our previous study of the HH 211 protostellar
system (Hirano et al. 2006; Palau et al. 2006; Lee et al 2007b, hereafter Paper I)
at unprecedented resolution up to $\sim$ \arcsa{0}{2}.
This system is located in the IC 348 complex of Perseus, with
an average distance of $\sim$ 280 pc \citep{Enoch2006,Lada2006}.
The central source is
a young, low-mass, and low-luminosity
($\sim 3.6$ \solarlum{}, corrected for the new distance) Class 0 
protostar with $T_{bol} \sim$ 31 K \citep{Froebrich2003}. 
It is seen with a highly collimated knotty jet in \H2{} \citep{McCaughrean1994}, CO
\citep{Gueth1999}, and SiO \citep{Hirano2006,Palau2006,Lee2007HH211},
driving a collimated outflow  \citep{McCaughrean1994,Gueth1999}.
It is also seen with a rotating envelope in ammonia \citep{Wiseman2001}.
Lying close to the plane of the sky 
\cite[$<$ 10\degree{},][]{Lee2007HH211},
this jet is one of the best candidates to search for jet rotation.
Previously, a velocity gradient was seen in SiO across the jet axis toward
the brightest knots of the jet at low spatial and velocity resolutions and
thought to be from jet rotation \citep{Lee2007HH211}.  However, since the
sense of rotation of the ammonia envelope is now found to be opposite to
that stated in \citet{Lee2007HH211} (J. Wiseman 2009, private communication;
H. Arce 2009, private communication), the velocity gradient found in
\citet{Lee2007HH211} is unlikely to
be from jet rotation. Here, we search for a velocity gradient that is
consistent with the sense of rotation of the ammonia envelope, with better
resolved shock structures and kinematics at $\sim$ 4 times better angular
and $\sim$ 3 times better velocity resolutions.
We also search
in 850 \micron{} continuum and \HCOP{} for a compact accretion disk argued to have
formed around the protostar \citep{Lee2007HH211}.
We also study in \HCOP{} a low-speed outflow that may carry away part of the
excess angular momentum from the envelope-disk.

\section{Observations}\label{sec:obs}

Observations toward the HH 211 protostellar system were carried out with the
SMA on 2008 Jan 23 in the extended configuration and on 2008 August 18 in
the very extended configuration. SiO ($J=8-7$), CO ($J=3-2$), SO (\SOt{}),
and \HCOP{} ($J = 4-3$) lines were observed simultaneously with 850
\micron{} continuum using the 345 GHz band receivers.  In this paper, we
only present the results in continuum, \HCOP{}, and SiO. The results in CO
and SO will be presented in a future publication.  The receivers have two
sidebands, lower and upper, covering the frequency range from 345.5 to 347.5
and from 355.5 to 357.5 GHz, respectively.  Combining the line-free portions
of the two sidebands results in a total continuum bandwidth of $\sim$ 3.7
GHz centered at $\sim$ 352 GHz (or $\lambda \sim$ 850 \micron{}) for the
continuum. The baselines have projected lengths ranging from $\sim$ 35 to
500 m. The primary beam has a size of $\sim$ \arcs{35} and one pointing was
used to map the central region of this system. The correlator was setup to have a
velocity resolution of $\sim$ 0.175 \vkm{} for the SiO and \HCOP{} lines.

The visibility data were calibrated with the MIR package,
with quasar 3C454.3 as a passband calibrator, and
quasars 3C84 and J0336+323 as gain calibrators.
The dwarf planet Ceres and the star-forming region MWC 349 were used as flux calibrators
in the extended and very extended configurations, respectively.
The flux uncertainty is estimated to be $\sim$ 15\%.
The calibrated visibility data were imaged with the MIRIAD package.
The dirty maps that were produced from the calibrated visibility data
were CLEANed using the Steer clean method,
producing the CLEAN component maps.
The final maps were obtained by restoring the CLEAN component
maps with a synthesized 
(Gaussian) beam fitted to the main lobe of the dirty beam. 
With natural weighting, the synthesized beam has 
a size of \arcsa{0}{46}$\times$\arcsa{0}{36} 
at a position angle (P.A.) of $\sim$ 70\degree{}.
The rms noise level is $\sim$ 0.09 \Jyb{} in the
\HCOP{} channel maps with a velocity resolution of 0.175 \vkm{},
$\sim$ 0.08 \mJyb{} in SiO channel maps with a velocity resolution of 0.35
\vkm{}, and $\sim$ 1.4 \mJyb{} in the continuum map.
Super-uniform weightings are also used to achieve
higher angular resolutions of up to \arcsa{0}{2}$\times$\arcsa{0}{15}.
The absolute position accuracy is estimated to be $\sim$ \arcsa{0}{03}.

\section{Results}

In the following, the systemic velocity in the HH 211 system is assumed to
be $9.2$ \vkm{} LSR, as derived from the optically thin line of \aHCOP{}
\citep{Gueth1999}.

\subsection{352 GHz Continuum Emission} \label{sec:cont}

\begin{figure} [!hbp]
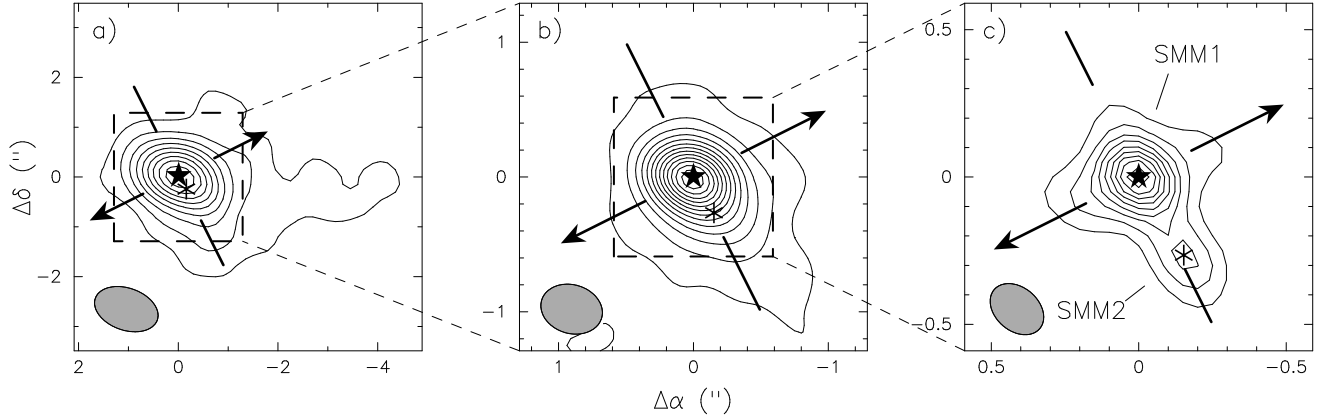
 
\centering \putfigs{0.7}{270}{f1.ps}
\figcaption[] 
{Continuum maps. The star and asterisk mark the positions of
the continuum sources SMM1 and SMM2, respectively. The arrows indicate the directions
of the western and eastern components of the jet. The thick lines indicate
the equatorial plane perpendicular to the jet axis. \tlabel{a} 342 GHz
continuum map at \arcsa{1}{28}$\times$\arcsa{0}{84} resolution from
\citet{Lee2007HH211}. The contours go from 10\% to 90\% of the peak value,
which is 155 \mJyb{} (1.51 K). \tlabel{b} 352 GHz continuum map made with natural
weighting, showing the inner part of the continuum emission. The bean is
\arcsa{0}{46}$\times$\arcsa{0}{36} with a P.A. of $\sim$ 73\degree{}.
Contour spacing is 8.4 \mJyb{} (0.5 K) with the first contour at 4.2 \mJyb{}
(0.25 K).
\tlabel{c} 352 GHz continuum map made with super-uniform weighting, showing
the innermost part of the continuum emission.  The beam is
\arcsa{0}{20}$\times$\arcsa{0}{15} with a P.A. of $\sim$ 49\degree{}.
Contour spacing is 6.4 \mJyb{} (2.1 K),
with the first contour at 6.4 \mJyb{} (2.1 K). 
\label{fig:cont}} \end{figure}

Previously, continuum emission was mapped around the source at 342 GHz
at $\sim$ \arcs{1} resolution, showing a dense flattened envelope in the
equatorial plane perpendicular to the jet axis and some (swept-up) envelope
material around the western outflow lobe \cite[Fig. \ref{fig:cont}a
and][]{Lee2007HH211}. The flattened envelope is asymmetric, extending more
to the southwest than to the northeast. Continuum emission is detected here
at 352 GHz at the higher resolution of $\sim$ \arcsa{0}{4} (Fig.
\ref{fig:cont}b), tracing mainly the inner part of the flattened envelope,
as the extended envelope emission is resolved out.
It has a total (integrated)
flux density of 0.22$\pm0.04$ Jy, a half of that seen in Figure \ref{fig:cont}a,
which is 0.44$\pm0.10$ Jy \citep{Lee2007HH211}.
Note that the inner part of the flattened envelope that extends $\sim$
\arcsa{1}{2} to the southwest is shifted slightly to the west
from the equatorial plane.

A compact continuum source, SMM1, is seen at the center of the jet with a
peak position $\alpha_{(2000)}=\ra{03}{43}{56}{804}$,
$\delta_{(2000)}=\dec{32}{00}{50}{27}$, as we zoom into the center at the
highest available resolution of $\sim$ \arcsa{0}{2}$\times$\arcsa{0}{15}
(Fig. \ref{fig:cont}c). 
It is seen with faint emission extending $\sim$  
\arcsa{0}{3} (84 AU)
to the northeast and southwest along the equatorial plane that
traces the innermost part of the flattened envelope or disk.
It is also seen with faint emission extending with a similar distance
to the northwest and southeast along the jet axis that may trace the jet
beam itself near the launching point.
Its peak position is well coincident with that
found at 43.3 GHz (or $\lambda= 7$ mm) at a similar resolution
\citep{Avila2001} and is thus considered as the position of the embedded
protostar driving the HH 211 jet. It is spatially unresolved and thus has
a deconvolved size smaller than a half of the beam size, or
$<$ \arcsa{0}{1} (28 AU) along the equatorial plane, as found at 43.3 GHz
\citep{Avila2001}. It has a flux
density of $\sim$ 80$\pm20$ mJy (integrated over a radius of \arcsa{0}{1})
and thus may have a true brightness temperature $>$ 80 K. The spectral index
$\alpha$ with $F_\nu\propto \nu^\alpha$ can be estimated with this flux and
the flux at 43.3 GHz that was measured for the similar region at a similar
resolution. With the flux of 2.7$\pm$0.6 mJy at 43.3 GHz \citep{Avila2001},
the spectral index is found to be $\alpha \sim$ 1.6, smaller than that for an optically
thick thermal emission, for which $\alpha=2$. It could be because
significant fraction of the flux at 43.3 GHz is from free-free emission of
an ionized gas or jet as in HH 111 \citep{Rodriguez2008}. If that is the
case, SMM1 could trace a warm optically thick (accretion) disk around the
embedded protostar, as suggested in \citet{Lee2007HH211}.
It seems to have a dust temperature $>$ 80 K, as implied from the true
brightness temperature.
Assuming that SMM1 has a dust temperature of 80 K and
its emission is optically thin, it has a (gas and dust) mass of only
1.1$\times10^{-3}\times2.84^\beta$ \solarmass{} or 1$-$3 \Jupmass{}, 
where $\beta$ is the dust opacity index ranging from 0 to 1. 
Here, a mass opacity
$\kappa_\nu = 0.1 (\nu/10^{12} \textrm{\scriptsize Hz})^\beta$
cm$^2$ g$^{-1}$ \citep{Beckwith1990} is assumed for the dust.
The mass can be lower, e.g., by $\sim$ 40\% if the
dust temperature is higher by 50\%.
In addition, since the emission is
likely to be optically thick, the mass here is only a lower limit.

A secondary continuum source, SMM2, is seen with a
peak position $\alpha_{(2000)}=\ra{03}{43}{56}{792}$,
$\delta_{(2000)}=\dec{32}{00}{50}{00}$
in the equatorial plane at $\sim$
\arcsa{0}{3} (or $\sim$ 84 AU ) to the southwest of SMM1, with S/N $\sim$
6. It has a flux density of $\sim$ 25$\pm10$ mJy and a deconvolved size of $<$
\arcsa{0}{1} (28 AU), leading to a true brightness temperature and thus a
dust temperature $>$ 25 K.
Assuming that the emission is optically thin with a dust
temperature of 25 K, it has a mass of 1.5$\times10^{-3}\times2.84^\beta$
\solarmass{} or $1.5-4$ \Jupmass{}, with $\beta$ ranging from 0 to 1.
Again, the mass can be lower, e.g., by $\sim$ 40\% if the
dust temperature is higher by 50\%.

\subsection{\HCOP{} emission}\label{sec:HCOP}

\begin{figure} [!hbp]
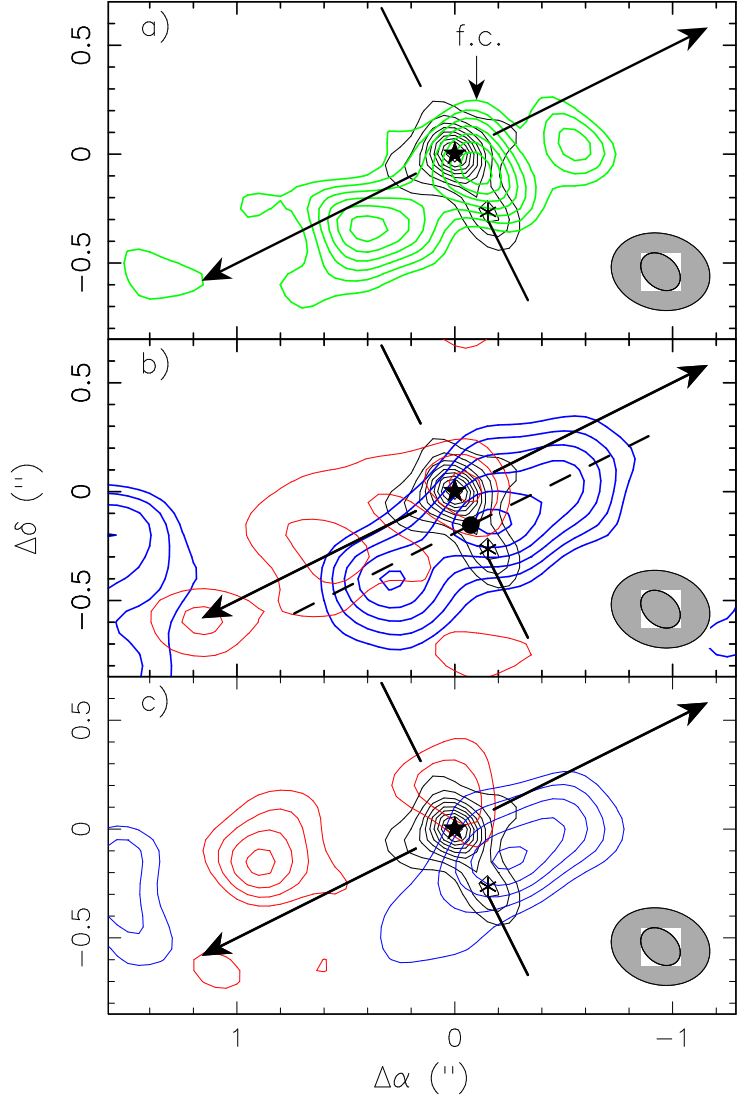

\centering
\putfigs{0.85}{270}{f2.ps}
\figcaption[]
{\HCOP{} emission on top of the continuum emission (black contours, as shown
in Fig. \ref{fig:cont}c).
The star, asterisk, arrows and solid lines have the same meanings as in
Fig. \ref{fig:cont}.
\tlabel{a} Green contours are the \HCOP{} emission integrated over
3.2 \vkm{} from $-$1.6 to 1.6 \vkm{} w.r.t. the systemic.
Contour spacing is 0.07 \Jybk{}, with the first contour at 0.14 \Jybk{}.
Here, f.c. means flattened condensation.
\tlabel{b} Blueshifted and redshifted emission, integrated from $-$1.6 to
$-$0.4 \vkm{} and from 0.4 to 1.6 \vkm{} w.r.t. the systemic, respectively.
Contour spacing is 0.09 \Jybk{}, with the first contour at 0.09 \Jybk{}.
The dashed line and the dot indicate the cut direction and cut center 
used for Figure \ref{fig:pvhcop}b.
\tlabel{c} Channel maps centered at $-$1 (blue) and 1 (red) \vkm{} from the systemic.
Contour spacing is 0.09 \Jyb{} (5.4 K), 
with the first contour at 0.18 \Jyb{} (10.8 K).
The synthesized beams are \arcsa{0}{20}$\times$\arcsa{0}{15}
for the continuum and \arcsa{0}{46}$\times$\arcsa{0}{35} for the \HCOP{}.
\label{fig:hcop}}
\end{figure}

\begin{figure} [!hbp]
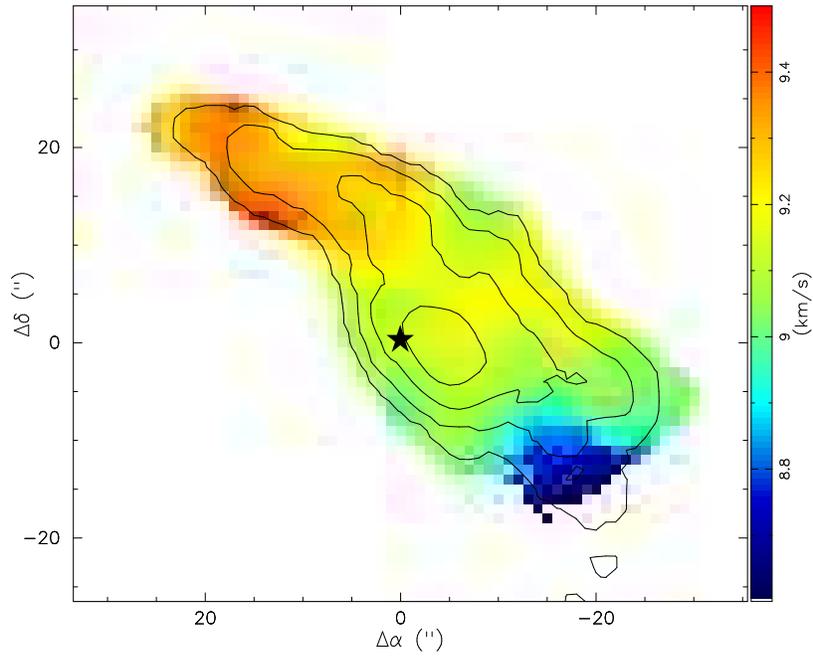

\centering
\putfigs{0.5}{270}{f3.ps}
\figcaption[]
{Velocity field (color shades) of the NH$_3$ envelope with
the integrated intensities (contours), as imaged with the VLA (Wiseman et al. in prep). 
A velocity gradient, with the redshifted emission in the northeast and blueshifted
emission in the southwest, is seen across the flattened
NH$_3$ envelope, as an evidence of rotation around the source SMM1 (as indicated
with a star).
\label{fig:nh3}}
\end{figure}

A flattened condensation is seen in \HCOP{} near SMM1 with a size of
$\sim$ \arcsa{0}{5}, slightly elongated along the equatorial plane (Fig.
\ref{fig:hcop}a). It peaks at $\sim$ \arcsa{0}{1} to the southwest of SMM1
and is shifted slightly to the northwest from the equatorial plane. Its
redshifted emission is in the northeast of its blueshifted emission
(Fig. \ref{fig:hcop}b), as seen in the large-scale
rotating ammonia envelope (Fig. \ref{fig:nh3}, Wiseman et al. in prep)\footnote{The sense of
rotation of the ammonia envelope (J. Wiseman 2006, private communication)
reported in Paper I was incorrect. Upon reexamining the ammonia data,
Wiseman (2009, private communication) reported a blueshifted NH$_3$ gas in
the southwest of the envelope, and a redshifted NH$_3$ gas in the northeast
of the envelope, which is opposite to that reported in \citet{Lee2007HH211}.
}, and thus it may trace a small-scale rotating envelope-disk
around SMM1. The center of rotation seems shifted slightly
away from SMM1 to the southwest in the direction of SMM2, judging from the
peak positions of the redshifted and blueshifted emission (Fig. \ref{fig:hcop}b). Since the
blueshifted emission is brighter than the redshifted emission, the emission
peak in the integrated map is shifted further to the southwest (Fig.
\ref{fig:hcop}a). Emission is also seen extending $\sim$ \arcsa{0}{7}
($\sim$ 200 AU) to the southeast and northwest around the jet axis,
perpendicular to the flattened condensation. 
This extended structure seems to rotate in the same direction as the
flattened condensation, with the blueshifted emission in the southwest and the
redshifted emission in the northeast (Fig. \ref{fig:hcop}b).
The redshifted emission is much weaker than the blueshifted
emission and even disappears in the northwest, probably because this
region is deeply embedded in a cold envelope that is expected to be
collapsing, so that most
of the redshifted emission is self-absorbed \citep{Evans1999}.

\begin{figure} [!hbp]
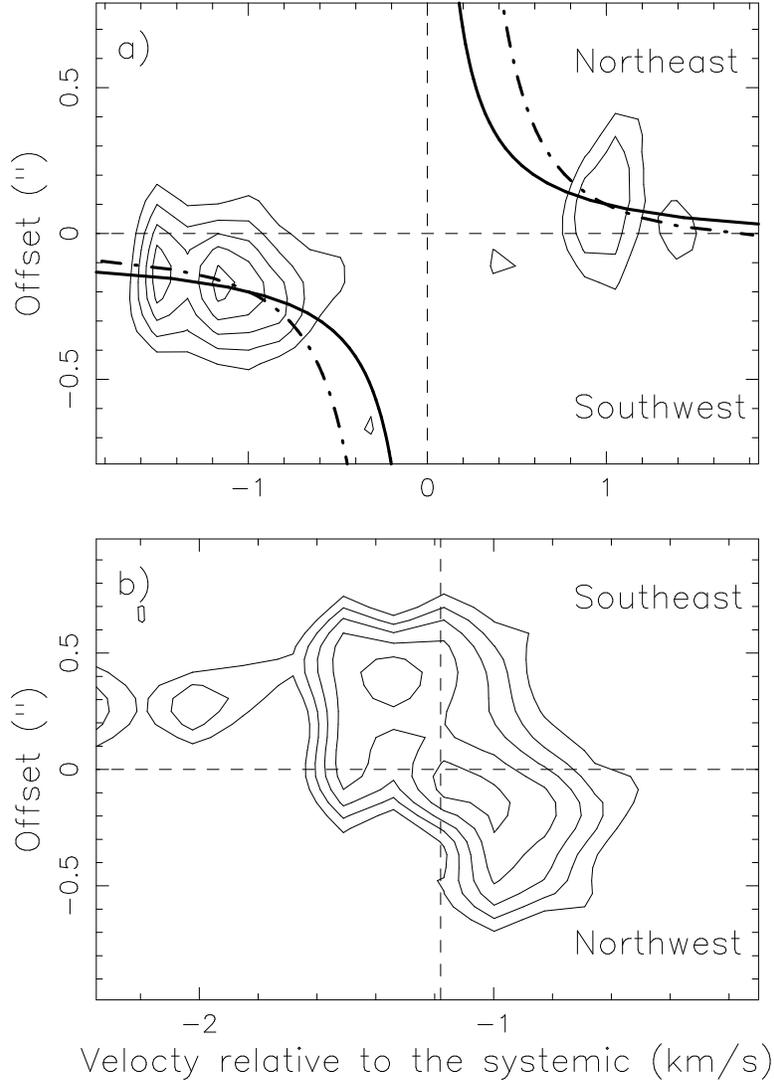

\centering
\putfigs{0.8}{270}{f4.ps}
\figcaption[]
{
\tlabel{a} Position-velocity diagram of \HCOP{} emission centered at SMM1
cut along the equatorial plane. Dashed curves are derived from Keplerian
rotation and solid curves are derived from a rotation law
$v_\textrm{\scriptsize rot}\propto r^{-1}$,
with the center of rotation shifted by
\arcsa{0}{05} to the southwest.
\tlabel{b} Position-velocity diagram of \HCOP{} emission centered at
\arcsa{0}{17} to
the southwest of SMM1, cut along the blueshifted part of the extended structure 
parallel to the jet axis, as indicated in Figure \ref{fig:hcop}b. 
The dashed vertical line indicates the rotation velocity at the cut
center, which is the base of the extended structure that has a rotation
veloicty of $\sim$ $-$1.2 \vkm{}.
In both panels, contour spacing is 0.09 \Jyb{} (5.4 K), 
with the first contour at 0.18 \Jyb{} (10.8 K).
\label{fig:pvhcop}}
\end{figure}

The position-velocity (PV) diagram cut across the flattened condensation along the
equatorial plane shows that the redshifted emission indeed is to the northeast and
the blueshifted emission is to the southwest of SMM1 with the velocity
increasing toward SMM1 (Fig. \ref{fig:pvhcop}a), as expected for a rotating
envelope-disk around SMM1.  This  envelope-disk 
is clearly seen in the redshifted emission at $\sim$1 \vkm{} w.r.t. the systemic (Fig.
\ref{fig:hcop}c), where it is not confused with the structure extended along
the jet axis.
Judging from the peak positions in the PV diagram at the highest
velocities at $\sim$ $\pm$1.5 \vkm{} w.r.t.  the systemic, the center of
rotation indeed seems shifted slightly to the southwest of SMM1, as
suggested in Figure \ref{fig:hcop}b.  Assuming that the shift of the center
is $\sim$ \arcsa{0}{05}, the velocity structure can be roughly fitted with a
Keplerian rotation with a central mass of $0.048\pm0.015$ \solarmass{}
(i.e., $\sim$ 50 Jupiter mass). This mass is consistent with that
derived from an evolution model by \citet{Froebrich2003}, which was found to
be 0.06 \solarmass{}. Therefore, the condensation may trace a rotationally
supported disk.  This disk seems to be truncated around SMM2 with a radius
of $\sim$ 80 AU (Figs. \ref{fig:hcop}a \& \ref{fig:pvhcop}a). The
compact continuum source SMM1 may trace the inner part of this disk. Note
that, since the velocity can be fitted as well by a rotation law 
with $v_\textrm{\scriptsize rot}\propto r^{-1}$, 
this condensation could also trace the inner part of a
non-rotationally supported flattened envelope \cite[see, e.g.,][]{Allen2003}.
We can not determine which rotation law can better fit our
observations.

\leftblank{
Assuming that the condensation has an excitation temperature
of 50 K, that the emission is optically thin and that the \HCOP{} abundance
is 2$\times10^{-9}$ \cite[highly uncertain, see][]{Girart2000}, the peak has a column density of
$1.4\times10^{22}$ \cms{} of molecular hydrogen. With a length along the
line of the sight assumed to be $\sim$
\arcsa{0}{5}, the density is estimated to be $\sim$ $7\times10^{6}$ \cmc{}, close to the
critical density. The mass of the condensation is $\sim$ 6$\times10^{-5}$
\solarmass{}.}


The structure extended to the southeast and northwest along the jet
axis perpendicular to the flattened condensation may have an outflow motion along the jet
axis, in addition to the rotation motion around the jet axis. The
blueshifted part of this structure is bright and thus can be used to check
this possibility, with a PV diagram
cut parallel to the jet axis (as indicated in Fig. \ref{fig:hcop}b).
This blueshifted part has a base in the flattened condensation at $\sim$
\arcsa{0}{17} in the southwest of SMM1, which has a rotation velocity of $\sim$
$-$1.2 \vkm{} (see Fig. \ref{fig:hcop}a).
Its emission
is shifted by this rotation velocity at the base
(Fig. \ref{fig:pvhcop}b), as expected if it is rotating
around the jet axis like the flattened condensation,
as suggested in Figure \ref{fig:hcop}b.
In addition, w.r.t. that at the base,
its emission is mainly slightly blueshifted (by $\sim$ $-$0.2 \vkm{}) in the southeast
and redshifted (by $\sim$ 0.2 \vkm{}) in the northwest, indicating that
it also has an outflow motion along
the jet axis, with the approaching side in the southeast and
the receding side in the northwest, as seen for the jet (Fig. \ref{fig:siojet}).
The projected outflow velocity, with a mean value of $\sim$ 0.2 \vkm{},
is small compared to the rotation velocity
because the outflow is almost in the plane of the sky.

\subsection{SiO Jet}

\subsubsection{Morphology}

\begin{figure} [!hbp]
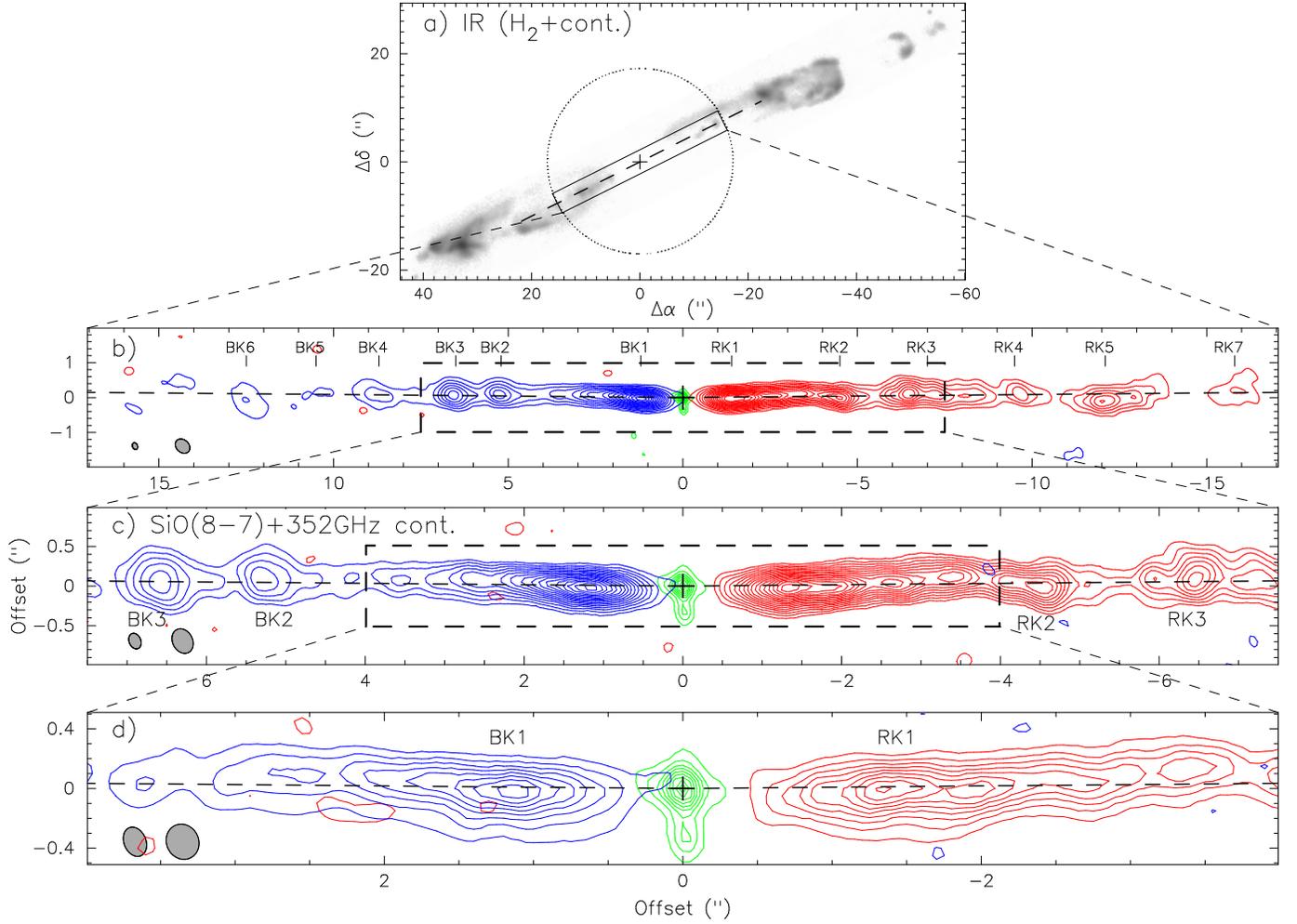

\centering
\putfigs{0.75}{270}{f5.ps}
\figcaption[]
{
\tlabel{a} The IR image from \citet{Hirano2006}.
The dotted line outlines our observed region.
The dashed line indicates the jet axis.
The cross marks the position of SMM1.
\tlabel{b}$-$\tlabel{d} 352 GHz continuum contours (green, as in Fig. \ref{fig:cont}c)
and redshifted (red) and blueshifted (blue) SiO contours.
The images are rotated by 26.6\degree{} clockwise.
The western and eastern
components of the jet axis are seen bent by $\sim$ 0.5\degree{} to the
north, as in \citet{Lee2007HH211}.
The cross marks the position of SMM1.
The knots have the same notations as in \citet{Lee2007HH211}.
The redshifted emission is integrated from 9.2 to 47.5 \vkm{}.
The blueshifted emission is integrated from -21.2 to 9.2 \vkm{}.
The synthesized beams are 
\arcsa{0}{46}$\times$\arcsa{0}{36} in \tlabel{b}, 
\arcsa{0}{35}$\times$\arcsa{0}{25} in \tlabel{c}, and
\arcsa{0}{24}$\times$\arcsa{0}{22} in \tlabel{d}.
First contours are 1 \Jybk{} in \tlabel{b}
and \tlabel{d}, and 0.84 \Jybk{} in \tlabel{c}.
Contour spacings are 1.5 \Jybk{} in \tlabel{b}
and \tlabel{d}, and 0.84 \Jybk{} in \tlabel{c}.
\label{fig:siojet}
}
\end{figure}

As seen in \citet{Lee2007HH211},
the SiO jet consists of a chain of paired knots
on either side of the source and it has
a mean jet axis at P.A. of 116.1\degree{}$\pm0.5$\degree{} and
297.1\degree{}$\pm0.5$\degree{}, respectively, for the eastern and western
components  (Fig. \ref{fig:siojet}).
At high resolution,
the innermost pair of knots, BK1 and RK1, are now seen as two linear curvy
structures on either side of the source, connecting to the continuum
emission (Fig. \ref{fig:siojet}d).
Knot BK1 extends to
$\sim$ \arcs{0.3} (84 AU) to the source with a faint emission overlapping
with the continuum emission extended to the east (Fig. \ref{fig:siojet}d). 
Knot RK1 extends to $\sim$ \arcs{0.5} (140 AU) to the source and points to the
continuum emission extended to the west. 
These two knots are very narrow with a deconvolved size
(i.e., transverse width) of $<$ \arcsa{0}{15} ($\sim$ 40 AU) and seem to
consist of $\sim$ 4 smaller sub-knots 
as indicated by the emission peaks.
They are curved in the same direction, first slightly to the south, then to
the north, and then to the south (Fig. \ref{fig:siojet}d). 
This is the so-called ``C-shaped bending"
and it could be due to an orbital motion of the source in a binary system
\citep{Fendt1998}. Note that the jet may also be slightly precessing, since 
the continuum emission extended along the jet axis is not exactly aligned 
with the jet axis.
That the two
knots, BK1 and RK1, do not bend exactly at the same distance further
supports this possibility. However, the
dominant effect must be the ``C-shaped bending", as the jet has an axial symmetry
(bending), not a point symmetry (precession).

Knots BK2, BK3, RK3, and likely also RK2, which are located further away
from the source, are now seen as head-tail structures (Fig.
\ref{fig:siojet}c). The head structures are due to the sideways ejection
of the internal shocks and grow bigger further away. The trailing tails may
trace the (weakly shocked) material in the jet beam. Knots BK4
to BK6 and RK4 to RK7 trace the internal (bow) shocks further downstream
(Fig. \ref{fig:siojet}b).

\subsubsection{Proper motion}

Proper motion of the SiO jet can be estimated by comparing the peak
positions of the knots here with those seen at $\sim$ 3.6 yrs
earlier in \citet{Lee2007HH211}.  We first convolved our maps here to the
resolution of those obtained earlier and then aligned the maps with the
continuum peaks (Fig. \ref{fig:pmjet}).  In addition, we only measure for
those knots that actually consist of one single (sub-)knot. They are knots
BK2, BK3, BK4, BK6, RK4, and RK7. The proper motion is estimated to be
$\sim$ \arcsa{0}{13}$\pm$\arcsa{0}{04} per year, resulting in a transverse
velocity of $\sim 170\pm$60 \vkm{} for the jet.

\subsubsection{Kinematics along the jet axis}\label{sec:kinematics}

A PV diagram of the SiO emission
cut along the jet axis is used to study
the kinematics of the jet (Fig. \ref{fig:pvjet}).
The jet has a mean radial velocity of $-$6 and 28 \vkm{}
LSR (or $-15.2$ and 18.8 \vkm{} w.r.t. the systemic) for the
eastern and western components, respectively. These mean radial velocities,
combined with the transverse velocity of the jet, result in an inclination of
$\sim$ $-$5\degree{}$\pm$2\degree{} and 6\degree{}$\pm$2\degree{} to the
plane of the sky for the eastern
and western components, respectively, and
a jet velocity of $\sim$ 170$\pm$60 \vkm{}.

In knot RK1, four tiny sub-knots with a separation of $\sim$ \arcsa{0}{75}
(210 AU) are seen with a range of velocities. Although it is not
that clear, similar number of sub-knots may form in knot BK1 as well.  The
velocity range is the broadest at the first sub-knot with a FWZM of $\sim$
30 \vkm{} (Fig. \ref{fig:specsio}), and it decreases with the distance from
the source. 
As mentioned, knots BK2, BK3, RK3, and likely also RK2 all show a head-tail 
structure in morphology (see Fig. \ref{fig:siojet}c). The
mean velocity of their tails is higher than that of their heads (Fig.
\ref{fig:pvjet}),
suggesting that the jet material in the tails will eventually go into the
heads.
The heads of knots BK2 and BK3 are better resolved longitudinally, and
their velocity is seen decreasing with the distance toward the downstream,
suggesting that they are formed
as the fast jet material catches up with the slow jet
material. This is expected if they are
the internal shocks produced
by a semi-periodical variation in the jet velocity,
like the knots in the IRAS 04166+2706 jet \citep{Santiago2008}.
On the other
hand, knot RK4, which is further away, shows a velocity increasing with the distance toward the
downstream. This change of velocity gradient at a larger distance
is also expected because of thermal expansion of the 
internal shock along the jet axis
\cite[see Fig. 8 in][]{Lee2004}.

\subsubsection{Jet rotation}

We present in Figure \ref{fig:pvrot} the PV diagrams cut across the jet
axis centered at the peaks of knots BK1, BK2, BK3, RK1, RK2, and RK3, in
order to follow up our previous study of jet rotation
at higher spatial and velocity resolutions.
In \cite{Lee2007HH211}, the
beam was \arcsa{1}{28}$\times$\arcsa{0}{84} in size and elongated with a
high inclination ($\sim$ 45\degree{}) to the jet axis. The velocity
resolution was binned to 1.0 \vkm{} per channel. Here the beam is $\sim$ 3-4
times smaller with a size of \arcsa{0}{32}$\times$\arcsa{0}{25} and less
elongated with a smaller inclination ($\sim$ 30\degree{}) to the jet axis. 
The velocity resolution is $\sim$ 3 times better, binned to 0.35 \vkm{} per
channel.

As expected, the transverse width of the knots increases from BK1 to BK3 and
from RK1 to RK3, due to sideways ejection of the internal shocks 
(Fig. \ref{fig:pvrot}).
Unlike that seen in \citet{Lee2007HH211}, 
no clear velocity gradient is seen across these knots.
Previously, the knots were not resolved and the apparent
gradient could result from a velocity gradient originally along the jet
axis, as discussed in \citet{Lee2007HH211}. Similar apparent
gradient is also seen in the PV diagrams derived from our channel maps 
degraded to the low spatial
and velocity resolutions of \citet{Lee2007HH211}.
The apparent gradient seen in \citet{Lee2007HH211} was uncertain and here we
show that it was misleading.

In order to further search for jet rotation,
we zoom into the innermost pair of knots, BK1 and RK1,
at the highest available resolution of \arcsa{0}{24}$\times$\arcsa{0}{22}
(Fig. \ref{fig:pvrot_h}).
Here we focus only on the innermost pair of knots because
the knots further downstream are more affected by the sideways ejection and 
jet precession (see Fig. \ref{fig:pvrot}).
The PV structures are now better resolved and can be separated into
two components, one
blueshifted and one redshifted, with respect to $\sim$ -6.5 km/s for knot BK1 and 
$\sim$ 29.1 km/s for knot RK1 (Fig. \ref{fig:pvrot_h}).
These two velocities are close to the mean
velocities of the jet and can be considered as the velocity centroids of the
two knots.
For knot RK1, it is clear that the redshifted component is mainly in the northeast
and the blueshifted component is mainly in the southwest,
similar to that seen in the \HCOP{} envelope-disk.
For knot BK1,
the two components also show these position offsets
at low velocity shifts, even though they extend to opposite sides at
high-velocity shifts.
The velocity structures are complicated and could be due to sideways ejection or jet rotation or
both.
Since sideways ejection is mainly perpendicular to the jet axis, the
velocity shifts due to sideways ejection are expected to be insignificant at the jet
edges (i.e., boundaries).
Thus, jet rotation is better measured with the velocity shifts at the two jet edges 
around the velocity centroid, which can be given by
the two peaks at ($-$7.1\vkm{}, $-$\arcsa{0}{04}) and ($-$5.9 \vkm{}, \arcsa{0}{03}) 
for knot BK1,
and (28.7 \vkm{}, $-$\arcsa{0}{04}) and (29.4 \vkm{}, \arcsa{0}{03}) for
knot RK1 (Fig. \ref{fig:pvrot_h}). 
The velocity gradients defined by these peaks could be real since the emission at the
two velocities that define the gradients are indeed slightly on the opposite
sides of the jet axis (Fig. \ref{fig:rotmap}).
The mean velocity gradient, with a velocity of $\sim$ 0.5 \vkm{} at $\sim$ 10 AU
(\arcsa{0}{035}),
if arising from jet rotation, would result in a
mean angular momentum of $\sim$ 5 AU \vkm{}.
Since a small-scale jet
precession could introduce velocity shifts that mimic jet rotation
\citep{Cerqueira2006} and
the position shifts between the two peaks that define the gradients
are less than one third of the 
synthesized beam, this mean velocity gradient could only be considered as an upper
limit of the true gradient due to jet rotation.

\subsection{Temperature and Density}

With lower transition lines of SiO,
the kinetic temperature of the SiO emission
has been estimated to be $>$ 300-500 K for the innermost pair of knots
\citep{Hirano2006}.
As mentioned, these knots are spatially
unresolved in the direction perpendicular to the jet axis
with a deconvolved size $<$ \arcsa{0}{15}.
Since their brightness temperature has a peak of $\sim$ 100 K at
\arcsa{0}{32}$\times$\arcsa{0}{25} resolution (Fig. \ref{fig:specsio}),
their true brightness temperature could have a peak 
close to their kinetic temperature.
Therefore, the SiO emission could be optically thick
as in the case of HH 212 \citep{Cabrit2007,Lee2007HH212}. 
Assuming that the
SiO emission has a kinetic temperature of 500 K and is optically thin, and
that the SiO abundance relative to molecular hydrogen is
$10^{-6}-10^{-7}$ \citep{Nisini2002,Hirano2006},
the density toward these knots is 10$^7-10^8$ \cmc{}.
Note that for the optically thick case,
the density is expected to be close to the critical density of SiO J=8-7
transition, which is $\sim$ 10$^8$ \cmc{}.


\section{Discussion}

\subsection{SMM1: A protostar with a mass of a brown dwarf?} \label{sec:SMM1}


Since SMM1 is seen with a \HCOP{} condensation that can
be explained with a Keplerian disk, the protostar at the peak of
SMM1 may indeed currently have a mass of only $\sim$ 50 \Jupmass{},
which is a mass of a brown dwarf.
The disk accretion rate can be estimated assuming that the bolometric luminosity
$\Lbol{}$ is mainly from the accretion. With 
$\Lbol \sim 3.6\, L_\odot$, a stellar mass of
$M_\ast \sim$ 50 \Jupmass{}, and a stellar radius of $R_\ast \sim (1-2) R_\odot$
\citep{Stahler1988,Machida2008},
the accretion rate $\dot{M}_a\sim \Lbol{} R_\ast/G
M_\ast \sim (2.5-5.0)\times 10^{-6}$ \smassrate{}.
If the accretion rate was the same in the past,
this protostar has an accretion age of only $\sim$ $(2-1)\times10^4$ yrs. 
If this protostar continues to accrete mass at this rate from the 
envelope (Fig. \ref{fig:cont}a) that has a mass of $\sim$ 50 \Jupmass{}
\citep{Lee2007HH211}, it
will start burning hydrogen in $\lesssim10^4$ yrs.

This protostar, if indeed with such a low mass at such a young age, could be
the lowest mass, youngest source known to have a rotating disk, a
high-speed collimated jet, and a collimated outflow. Among the youngest low
luminosity objects known to date very few are clearly in the Class 0 stage
and driving (high-speed) collimated outflows, with possibly IRAM 04191
\citep{Andre1999,Lee2005IRAM} and IRAS 04166+2706
\citep{Tafalla2004,Santiago2008} being the most remarkable cases. However, neither
IRAM 04191 nor IRAS 04166+2706 have been reported to have a rotating
structure at the disk scale ($\lesssim$ 100 AU). Other brown dwarf
candidates associated with disks (inferred from the spectral energy
distribution) and driving outflows are all in more evolved evolutionary
stages (Class I/II/III) \citep{White2004,Bouy2008,Phan-Bao2008}. Thus, the
HH 211 system, revealing a substellar object (at the moment) with a rotating
envelope-disk in the Class 0 phase and driving a spectacular outflow
suggests that brown dwarfs and low mass stars form in a similar way, and
that whether they will become low-mass stars or brown dwarfs depends on the
amount of material that can be accreted still in the envelope.

\subsection{SMM2: A binary companion with a planetary mass?}

The nature of the continuum source SMM2 is uncertain. It may trace an
asymmetry in the inner envelope around SMM1. It may also trace a dusty
envelope-disk around another embedded source. 
That the \HCOP{} envelope-disk seems to have a center of rotation
shifted slightly away from SMM1 to the southwest
in the direction of SMM2 and that the \HCOP{} envelope-disk
seems to be truncated around SMM2 both support this possibility. The center of
mass in the system, if assumed to be the same as the center of rotation, 
is $\sim$ \arcsa{0}{05} in the southwest of SMM1 (see \S \ref{sec:HCOP}) and thus $\sim$
\arcsa{0}{25} in the northeast of SMM2. 
Thus, the embedded source, if existed, would have a 
mass of $\sim$ 8 \Jupmass{} (a mass lower than that of a brown dwarf),
with the protostar in SMM1 now having a mass of $\sim$ 42 \Jupmass{}.
Therefore, it would
be a small binary companion surrounded by a dusty envelope-disk of 1.5$-$4
\Jupmass{} (see \S \ref{sec:cont}). 
Further submillimeter observations are needed to confirm this.
The small mass of this companion may
explain why there is no clear evidence of an outflow associated with SMM2. This
companion would have an orbital period of $\sim$ 3000 yrs and could be
responsible for the large-scale precession of the jet \citep{Eisloffel2003}.
This binary companion, however, may not be responsible for the C-shaped
bending that seems to have a much shorter period.
We speculate that the HH\,211 system may eventually evolve
into a binary system consisting of a very low-mass star and a brown dwarf.


\subsection{Outflowing gas from the envelope-disk?}

The \HCOP{} structure extending $\sim$ \arcsa{0}{7}
($\sim$ 200 AU) to the southeast and northwest around the SiO jet seems
to have an outflow motion along the jet axis, in addition to the rotation
motion around the jet axis.
It is seen within a wide-opening cavity with the western walls traced by the western
extensions in the continuum (Fig. \ref{fig:cont}a) and is thus different from the
large-scale molecular outflow driven by the fast-moving jet and wind \citep{Lee2000}.
Assuming that it has the same inclination as the jet, it has
a deprojected outflow velocity of $\sim$ 2.1 \vkm{}, about
1.8 times the rotation velocity
at the base of the outflow, which is $\sim-1.2$ \vkm{} (Fig.
\ref{fig:pvhcop}b). Thus, it can be considered as a low-speed outflow.
With a size of $\sim$ 200 AU, the outflow has a dynamical time $\sim$ 450
years.

Similar low-speed outflow has also been seen along the jet axis 
in other Class 0 sources, e.g., IRAM 04191
\citep{Lee2005IRAM} and HH 212 \citep{Lee2007HH212} in \HCOP{}, and even in
Class II sources, e.g., RNO 91 \citep{Lee2002} and CB 26
\citep{Launhardt2008} in CO.  Like the HH 211 outflow, the HH 212 and CB 26
outflows may also be rotating.
In the case of CB 26, the outflow 
may arise from the disk and play a significant role in dispersing 
the disk material in the late stage of star formation \citep{Launhardt2008}. 
Here in the case of HH 211 (and probably also HH 212), the outflow 
may arise from the envelope-disk.
It has a mean specific angular momentum of $\sim$ 50 AU
\vkm{}, with $\sim$ 1.2 \vkm{} at $\sim$ 45 AU (\arcsa{0}{17}), and may
carry away part of the extra angular momentum, if not all, allowing material to fall
toward the center in the early phase of star formation.
With an outflow velocity to rotation velocity ratio of $\sim$ 2,
the low-speed outflow could be driven by magneto-centrifugal force as that seen 
in from the rotating structures, such as the outer part
of the disk \citep{Pudritz2007} or the inner part of 
the pseudodisk (a non-rotationally supported flattened envelope) 
\cite[e.g.,][]{Allen2003}.


\subsection{Origin of SiO emission near the source}

SiO emission is absent near the source, but suddenly appears at $\sim$
\arcsa{0}{5} in knots RK1 and BK1 and peaks at $\sim$ \arcsa{1}{2} at their
first sub-knots.  As mentioned, these knots likely consist of about 4
sub-knots separated by \arcsa{0}{7} with a velocity range decreasing with
the distance from the source. Near the source,
this decrease in velocity range with the distance is expected if the sub-knots trace the
internal shocks produced by a semi-periodical variation in the jet velocity
[for the detailed velocity structures produced by such variation, see e.g.,
Fig. 2 in \citet{Suttner1997} or Fig. 8 in \citet{Lee2004}].
If this is the
case, a sub-knot is expected to be seen closer in at $\sim$ \arcsa{0}{5}
from the source. Indeed, the SiO emission is seen extending all the way to
$\sim$ \arcsa{0}{5}. No such sub-knot (emission peak) is seen there
probably because the shock there has just started to develop and is too weak
to produce an emission peak. Thus, the SiO emission of the innermost pair of
knots seems closely related to the shock enhancement of SiO in gas phase.

SiO abundance can be greatly enhanced in the shocks as a
consequence of grain sputtering or grain-grain collisions releasing
Si-bearing material into the gas phase, which reacts rapidly with O-bearing
species (e.g., O$_2$ and OH) to form SiO \citep{Schilke1997,Caselli1997}.
This shock enhancement mechanism, which requires a flow time $>$ 100 yrs
\citep{Gusdorf2008a}, seems too slow to explain
the SiO emission of the innermost pair of knots, which have a dynamical time $<$ 30 yrs
with a jet velocity of $\sim$ 170 \vkm{}, as in HH 212
\citep{Cabrit2007,Lee2008}.
For the innermost SiO emission at $\sim$ \arcsa{0}{5} (140 AU) from the
source, the dynamical time is only $\sim$ 4 yrs.
Therefore, at least for the innermost SiO emission,
it is possible that SiO was already formed on dust grain and then
released in gas phase in the shocks \citep{Gusdorf2008b}.

It is also possible that the decrease in velocity range with the distance
for the innermost
pair of knots is explained in the context of a X-wind model, in which the
jet is actually a dense part of a wide-angle {\it radial} wind
\citep{Shu1995,Shang1998}. For a simple comparison, we can assume the jet
has a constant transverse size of $\sim$ \arcsa{0}{15}, then the velocity
range is given by $\sim 170 \cdot (\arcsaq{0}{15}/d)$ \vkm{}, where $d$ is
the distance of the jet from the source. Therefore, the velocity range is
expected to be $\sim$ 26 \vkm{} at \arcs{1} and $\sim$ 9 \vkm{} at \arcs{3},
roughly similar to that seen in the PV diagrams (Fig. \ref{fig:pvjet}). In
this case, SiO of gas phase could have formed in the jet due to high gas
density in the jet \citep{Glassgold1991}, with a mass-loss rate $\sim$
$10^{-6}$ \smassrate{} \citep{Lee2007HH211}. Further work is needed to study
its production rate in order to explain why the SiO emission is not seen all
the way to the source.

\subsection{Launching radius of the jet}

As mentioned,
the mean velocity gradient seen across the jet axis toward the innermost pair of
knots could be considered as an upper limit of the true gradient due to jet
rotation.
If this is the case, assuming that the SiO
jet is a dense part of a magnetized wind launched from an accretion disk
by magneto-centrifugal force, we can estimate 
the upper limit of the launching radius of the jet. With a
central mass of $\sim$ 50 \Jupmass{}, a jet velocity of $\sim$ 170
\vkm{}, and a rotation velocity of $\sim$ 0.5 \vkm{} at $\sim$ 10 AU, 
the upper limit of the launching radius is estimated to be $\sim$ 0.014 AU
($\sim$ 3 $R_\odot$), following \citet{Anderson2003}.
At this radius, the
Keplerian velocity is $\sim$ 54 \vkm{}, resulting in a jet velocity to
Keplerian velocity ratio of $\sim$ 3, close to that expected for
a wind launched by magneto-centrifugal force
\cite[e.g.,][]{Shu1995,Pudritz2007}.


Velocity gradients have also been seen in another Class 0 source HH 212 
across the jet axis
toward its innermost pair of SiO knots, SN and SS, and may arise from jet
rotation as well \citep{Lee2008}.
Since knot SN is closer to the source and less affected by sideways ejection,
its velocity gradient, with $\sim$ 0.5 \vkm{} at $\sim$ 50 AU, can be 
considered as an upper limit of the true gradient due to jet rotation.
Since the mass of the central protostar is $\sim$ 0.15 \solarmass{}
\citep{Lee2006}, the jet velocity is expected to be higher than that of HH
211 and can be assumed to be 200 \vkm{}. Thus, 
the upper limit of the launching radius of the
SiO jet is found to be $\sim$ 0.05 AU for the HH 212 system.
At this radius, the Keplerian
velocity is $\sim$ 52 \vkm{}, resulting in a jet velocity to Keplerian
velocity ratio of $\sim$ 3.8, also close to that
expected for a wind launched by magneto-centrifugal force.

As a result, if the velocity gradients are really due to jet rotation,
the SiO jets of the two Class 0 sources could be a part of
a magnetized wind launched at $\sim$ 0.014-0.05 AU from the source, 
much closer than that of the \HCOP{} outflow, which seems to have an outflow
base at the flattened envelope-disk.
They could be
launched from the innermost edge of the disk near the corotation radius, as
predicted in the X-wind model \citep{Shu1995}.
They may carry away excess angular momentum from
the inner edge of the disk, allowing material to fall on to the protostar.
Note that, however, since the velocity gradients could only be
considered as upper limits of the true gradients due to jet rotation, the
jets could have no rotation and different launching mechanism.
Further observations at higher angular resolution are really
needed to check this possibility.


\section{Conclusions}

We have mapped the HH 211 protostellar system in 352 GHz continuum, SiO
($J=8-7$), and \HCOP{} ($J=4-3$) emission at up to $\sim$ \arcsa{0}{2}
resolution with the SMA. The main conclusions are the
following:
\begin{enumerate}
\item 

The continuum source is now resolved into two sources, SMM1 and SMM2, with a
separation of $\sim$ 84 AU.  SMM1 is seen at the center of the jet, probably
tracing a (inner) dusty disk around the protostar driving the jet. Its mass
is estimated to be $\sim$ 1--3 \Jupmass{}. SMM2 is seen to the southwest of
SMM1 and may trace a 1.5--4 \Jupmass{} envelope-disk around a small binary
companion.

\item 
A flattened envelope-disk is seen in \HCOP{} around SMM1 perpendicular to
the jet axis, with a radius of $\sim$ 80 AU. Its velocity structure is
consistent with a rotation motion and can be fitted with a Keplerian
law that yields a central mass of $\sim$ 50$\pm$15 \Jupmass{}. Note that
its velocity structure can also be fitted as well with a
rotation law $\vrot \propto r^{-1}$.

\item 

If the flattened envelope-disk is indeed in Keplerian rotation, then the
protostar that drives the jet could have a mass of $\sim$ 50$\pm$15
\Jupmass{}. Thus, the protostar could be the lowest mass source known to
have a collimated jet and a rotating flattened envelope-disk.

\item 

A small-scale ($\sim$ 200 AU) low-speed ($\sim$ 2 \vkm{}) outflow is seen in
\HCOP{} around the jet axis extended from the envelope-disk. It seems to
rotate in the same direction as the envelope-disk with a mean specific
angular momentum of $\sim$ 50 AU \vkm{}, and thus may carry away part of the
angular momentum from the envelope-disk.

\item 

The jet is seen in SiO originating from SMM1. The innermost pair of knots are
now seen with a ``C-shaped" bending and each consisting of $\sim$ 4 smaller
sub-knots. They are very narrow, with a transverse width of $\lesssim$ 40
AU. Knots further downstream have a head-tail structure, with the head
likely resulting from sideways ejection of the internal shocks and the tails
tracing the weakly shocked material in the jet beam. The kinematics
associated with most of the head-tail knots is consistent with the fast jet
material in the tails catching up with the slower material in the heads. The
knots likely result from the internal shocks produced by a semi-periodical
variation in the jet velocity.

\item 

We measure the proper motion of the SiO knots by comparing the data between
2008 and 2004, and obtain a transverse velocity of 170$\pm$60 \vkm{}.
The jet, with an inclination of $\sim$ 5\degree{} to the plane of the
sky, has a velocity the same as the transverse velocity.

\item 

The SiO associated with the innermost pair of knots could not be formed by
grain-sputtering or grain-grain collisions that release Si into the gas
phase. We suggest that the SiO could form directly on the grain surfaces and
be released to the gas phase in the shocks.

\item

A possible velocity gradient is seen consistently across the innermost pair
of SiO knots, consistent with the rotation sense of the \HCOP{}
envelope-disk and the large-scale ammonia envelope. If this gradient, with
$\sim$ 0.5 \vkm{} at $\sim$ 10 AU, is an upper limit of the true rotational
gradient of the jet, then the jet carries away a very small amount of
angular momentum of $\lesssim$ 5 AU \vkm{} and thus must be launched from
the very inner edge of the disk, as predicted in the X-wind model.
Further observations at higher resolution are really needed to confirm
if this gradient is really due to jet rotation.


\end{enumerate}

\acknowledgements
We thank the SMA staff for their efforts
in running and maintaining the array, and the anonymous referee for
the precious comments.
C.-F. Lee thanks Jennifer Wiseman
for reexamining her ammonia data, Mark Gurwell for the help with flux
calibration, Ronald Taam, Ruben Krasnopolsky, and
Mike Cai for fruitful conversations.


\begin{figure} [!hbp]
\centering
\putfigs{0.75}{270}{f6.ps}
\figcaption[]
{Redshifted and blueshifted components of the SiO jet at two different times at the 
resolution of \arcsa{1}{28}$\times$\arcsa{0}{84}.
\tlabel{a} 
2004 Sept 10 (in extended configuration) + 2004 Oct 4 (in compact
configuration). The maps are from \citet{Lee2007HH211}, with the same velocity
ranges as in Fig. \ref{fig:siojet}b. 
Contour spacing is 3 \Jybk{}, with the first contour at 6 \Jybk{}.
\tlabel{b} 2008 Jan 23 (in extended configuration) + 2008 Aug 18 (in very extended
configuration). The maps are obtained by convolving our maps in Fig. \ref{fig:siojet}b.
Contour spacing is 3 \Jybk{}, with the first contour at 2 \Jybk{}.
\label{fig:pmjet}
}
\end{figure}

\begin{figure} [!hbp]
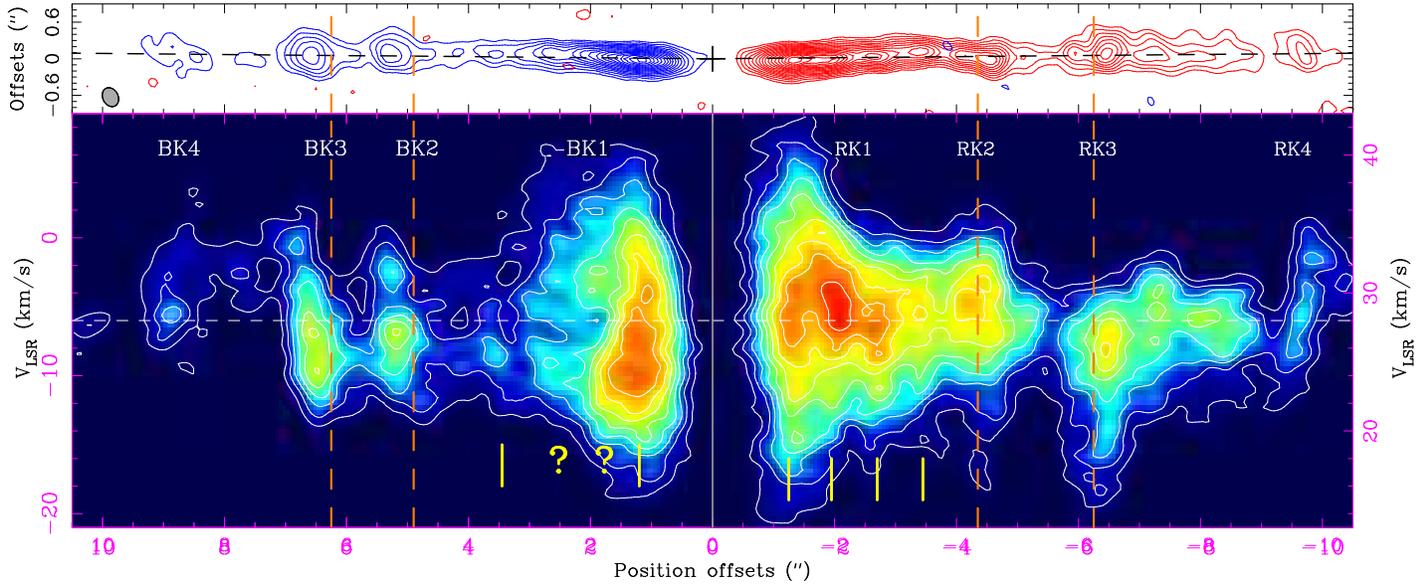

\centering
\putfigs{0.75}{270}{f7.ps}
\figcaption[]
{PV diagram of the SiO emission cut along the jet axis.
The orange dashed lines separate the heads and tails for knots BK2, BK3,
RK2, and RK3. The yellow lines mark the positions of the sub-knots in
knots BK1 and RK1. The question marks indicate the possible positions of the
two sub-knots in knot BK1. The systemic velocity is 9.2 \vkm{}. The white horizontal
dashed lines mark the mean velocities, which are $-$6 \vkm{}
on the blueshifted side and 28 \vkm{} on the redshifted side.
Contour spacing is 0.15 \Jyb{} (19 K) with the first contour at
0.15 \Jyb{} (19 K).
\label{fig:pvjet}
}
\end{figure}

\begin{figure} [!hbp]
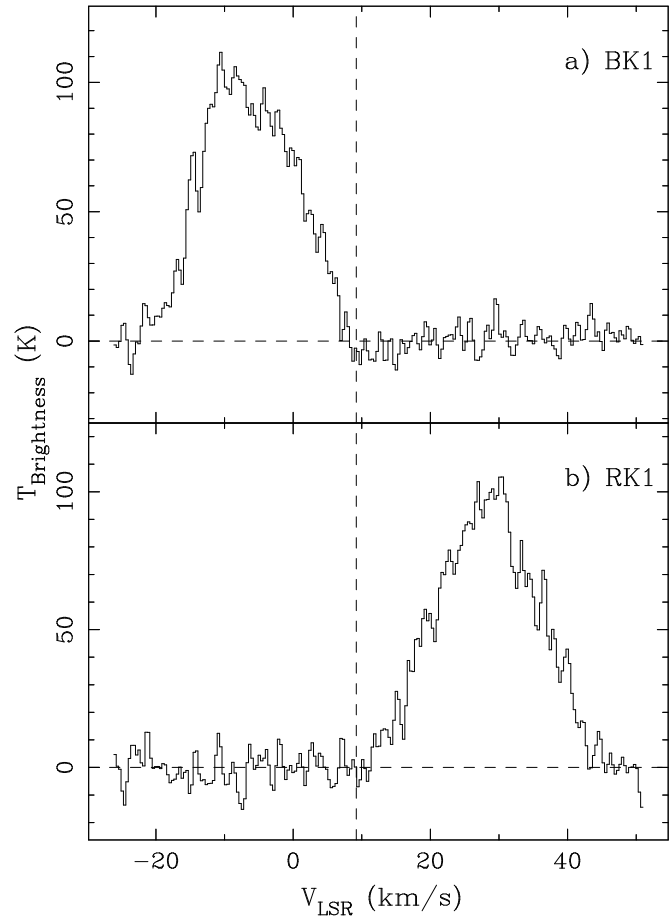

\centering
\putfigs{0.7}{270}{f8.ps}
\figcaption[]
{SiO spectra toward the first sub-knots of knots BK1 and RK1. Vertical dashed
lines indicate the systemic velocity.
\label{fig:specsio}
}
\end{figure}

\begin{figure} [!hbp]
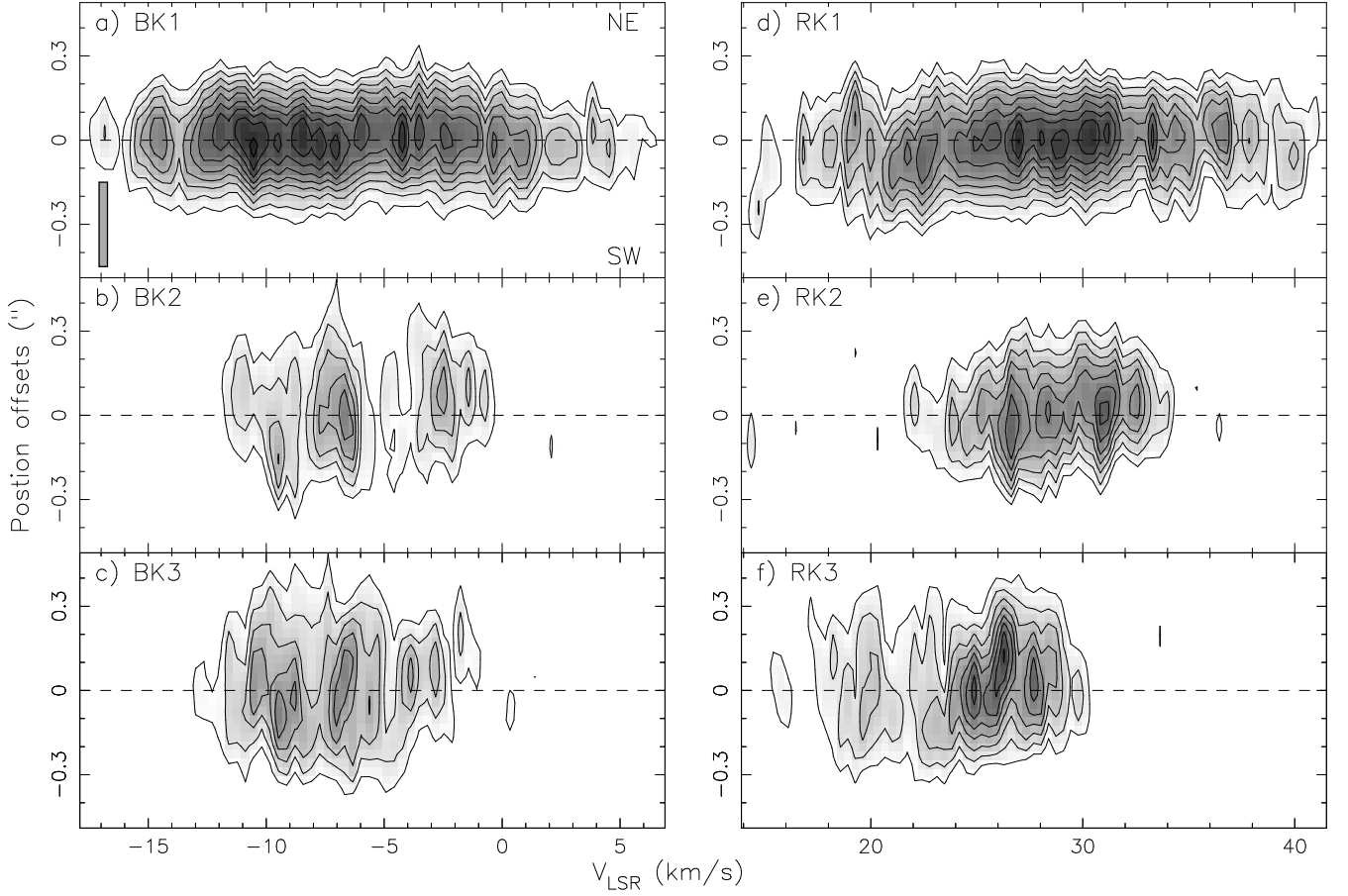

\centering
\putfigs{0.7}{270}{f9.ps}
\figcaption[]
{PV diagrams of the SiO emission cut across the jet axis centered at the peaks of
knots BK1, BK2, BK3, RK1, RK2 and RK3. 
The diagrams are derived from channel maps obtained with a beam of
\arcsa{0}{32}$\times$\arcsa{0}{25}. The cuts have a width of \arcsa{0}{1}.
Contour spacing is 0.085 \Jyb{} (10.76 K) 
with the first contour at 0.17 \Jyb{} (21.53 K).
The angular and velocity resolution is shown
in the lower left corner in \tlabel{a}.
\label{fig:pvrot}
}
\end{figure}

\begin{figure} [!hbp]
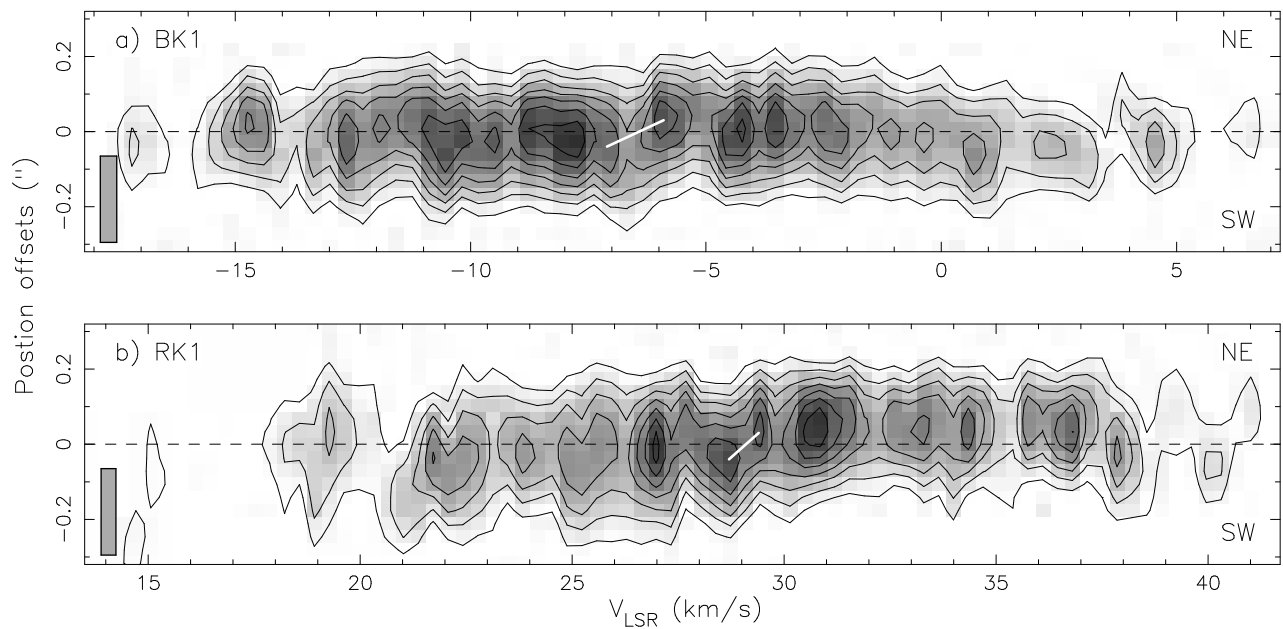

\centering
\putfigs{0.7}{270}{f10.ps}
\figcaption[]
{PV diagrams of the SiO emission cut across the jet axis centered at the peaks of
knots BK1 and RK1. The diagrams are derived from channel maps obtained with a beam of
\arcsa{0}{24}$\times$\arcsa{0}{22}.
The cuts have a width of \arcsa{0}{1}.
The white lines define the velocity gradients across the jet axis by
connecting the two peaks on the opposite sides.
Contour spacing is 0.085 \Jyb{} (16.31 K) 
with the first contour at 0.17 \Jyb{} (32.62 K).
The angular and velocity resolution is shown in the lower left corner.
\label{fig:pvrot_h}
}
\end{figure}

\begin{figure} [!hbp]
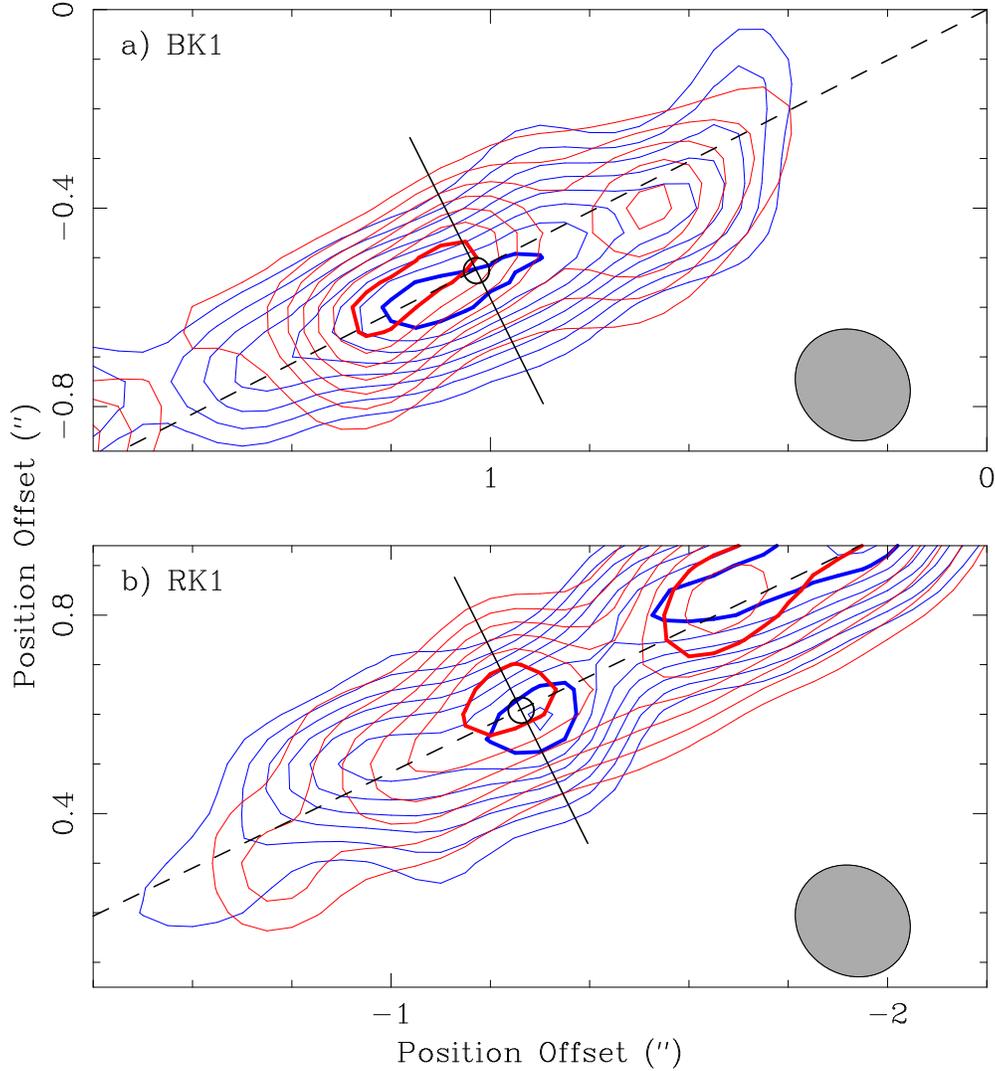

\centering
\putfigs{0.8}{270}{f11.ps}
\figcaption[]
{Maps 1 \vkm{} wide of the two line peaks, defining  
the velocity gradient shown in Fig. \ref{fig:pvrot_h}
for knots \tlabel{a} BK1 and \tlabel{b} RK1.
The circles and solid lines show the centers (where the emission peaks are in
the integrated map in Fig. \ref{fig:siojet}d) and the directions of the cuts used in the PV diagrams.
The dashed lines show the jet axes connected to the source.
Contour spacing is 0.075 \Jyb{} (14.63 K) with the first contour at
0.15 \Jyb{} (29.26 K). The two panels show that the redshifted emission and
blueshifted emission around the cut centers are indeed on the opposite sides
of the jet axis. Contours are highlighted to guide the eyes.
The beam has a size of \arcsa{0}{24}$\times$\arcsa{0}{22}.
\label{fig:rotmap}
}
\end{figure}

\end{document}